\documentstyle[amsfonts,12pt,sw20lart]{article}
\input{tcilatex}
\begin{document}

\title{{\bf A deformation of Hermite polynomials }}
\author{{\bf Mekhfi.M } \\
%EndAName
\\
{\normalsize Int}$^{^{\prime }}${\normalsize Centre For Theoretical
Physics,34100Trieste,ITALY}}
\date{2000 oct 1}
\maketitle

\begin{abstract}
We propose and study the properties of a set of polynomials $M_{n\alpha ,H\
}^{s}(z)$ , $C_{n\alpha ,H}^{s}(z)$ $W_{n\alpha ,H}^{s}(z)$ with $n,s\in \ N$
$;\alpha =\pm 1;$and where $H$ stands for Hermite ; the ''root '' polynomial
.These polynomials are obtained from a deformation of Hermite polynomials $%
H_{n}(z).$The structure underlying the deformation seems quite general and
not only restricted to Hermite polynomials.



{\footnotesize mekhfi@hotmail.com}
\end{abstract}

\begin{center}
\smallskip \newpage {}
\end{center}

\section{Introduction}

The concept of deformed exterior derivative $d_{\lambda }$ has been
introduced by Witten \cite{mohsine} who applied it successfully to Morse
theory .In this and in recent papers \cite{wissale} we make a different use
of the concept .More precisely ,we investigate special functions and
orthogonal polynomials in the context of deformed exterior derivative .To be
self contained we would like to first recall the definition of $d_{\lambda
}, $ and briefly mention how this , has been used to investigate Morse
theory and the way we used the formalism to unify some special functions and
how to apply it to orthogonal polynomials.

Let ${\cal R}$ be a Riemannian manifold of dimension n .Let $d$ be the usual
exterior derivative acting on p-forms.Let $V$ be a smooth function $V{\cal \ 
}$\thinspace $\,{\cal R}\rightarrow R\ or\ C$ and $\lambda $\ a real
number.Define

\begin{equation}
d_{\lambda }[V]=e^{-\lambda V}d\ e^{\lambda V}  \label{eq:1}
\end{equation}

The crucial remark made by Witten is that Betti's numbers $B_{p}(\lambda )$
of the manifold $M$ i.e.the number of linearly independent p-forms which
obey $d_{\lambda }\Psi =0$ but cannot be written as $\Psi =d_{\lambda }\
\chi $ for any $\chi $ is independent of $\lambda $ and therefore is equal
to the usual Betti's number $B_{p}.$ It then follows from standard arguments
that the number of zero eigenvalues of the Hamiltonian $H_{\lambda
}=d_{\lambda }d_{\lambda }^{*}+d_{\lambda }^{*}d_{\lambda }$ acting on
p-forms is just as at $\lambda =0$ equal to $B_{p}.$ This is useful because
it can be shown that the spectrum of $H_{\lambda }$simplifies dramatically
for large $\lambda .$This is the way the author was able to place upper
bounds on the $B_{p}$ in terms of the critical points of $V$ by studying the
spectrum of $H_{\lambda }$ for large $\lambda $.

In \cite{wissale} we investigated the simpler system

\begin{equation}
H_{\lambda }=d_{\lambda }  \label{eq:2}
\end{equation}

and showed that it gave information on the index structure of Bessel
functions.We note at once that the system \ref{eq:2} is topological in the
sense that $d_{\lambda }$ like $d$ is defined purely in terms of
differential topology without choosing a metric in $M$ .The investigation of 
$H_{\lambda }$ we made differs from that of reference \cite{mohsine} in
several aspects.First we restrict ourselves to 0-forms i.e. to functions on $%
M$ .Second we enhance the function V to act as an operator (non trivial
operator ).This will not spoil the property $d_{\lambda }^{2}=0$.Third $V$
is no longer arbitrary but should be properly chosen, and its form depends
on the problem considered.Finaly we are interested in nonvanishing
eigenstates of the system $H_{\lambda }$ and not on zero eigenstates and
their number( given by $B_{p}$'s) . More precisely we are interested in the
relationship between nonvanishing eigenstates of $d_{\lambda }$ and $d$.The
functions on $M$ we are interested in are the generating functions $\Phi
(z,t)=\sum_{n\in N\ or\ Z}\phi _{n}(z)t^{n}$ of a class of special functions
or orthogonal polynomials.In the process of deformation ,that is , in going
from $d$ to $d_{\lambda }$ the parameter $t$ is kept fixed i.e. $d=dz\frac{%
\partial }{\partial \ z}$ .Most special functions and orthogonal polynomials 
$\phi _{n}(z)$ obey recursion formulas involving the derivative with respect
to the variable z and therefore their generating functions can be made
eigenstates of $d$.It then follows from \ref{eq:1} that the function

\[
e^{\lambda V}\Phi 
\]

is presumably an eigenstate of $d_{\lambda }$ $,$ that is a deformation of $%
\Phi $ and could represent new special functions or new polynomials (wee
will see that orthogonality is not guaranteed ).We have shown in \cite
{wissale} that with a judicious choice of $V$ we indeed map integer order
Bessel functions into real orders .The appropriate form of the operator $V\ $%
is

\[
V_{\alpha }(z)=\sum_{m\in Z/0}\alpha ^{m}\frac{\prod_{m\alpha }(z)}{m} 
\]

where $\alpha =\pm 1$ and where the operator $\Pi $ is the one appearing in
the recursion formula involving the derivative with respect to z.For the
case of (reduced )Bessel functions $\frac{J_{n}}{z^{\alpha n}}\ $this
operator has the form $\prod_{m\alpha }(z)\ =(-\alpha )^{m}\frac{d^{m}}{(\
zdz\ )^{m}}$ and acts as $\prod_{m\alpha }(z)\ \frac{J_{n}}{z^{\alpha n}}=%
\frac{J_{n+\alpha m}}{z^{\alpha n+m}}$ The result of the deformation or
mapping \footnote{$d_{|m|}=\frac{d}{zdz}.\,\frac{d}{zdz}\cdots \cdots \frac{d%
}{zdz}.$ {\footnotesize and }$d_{-\ |m|}=\int zdz.\,\int
zdz..............\int zdz${\footnotesize \ where we extend the index m to
negative values by introducing the symbol }$\int dz${\footnotesize \ to
denote a ``truncated '' primitive i.e. in defining the integral we omit the
constant of integration }$\int \frac{df}{dz}dz=f${\footnotesize \ }} is

\[
\frac{J_{n+\alpha \lambda }}{z^{\alpha n+\lambda }}=\exp [-\lambda
\sum_{m\in Z/0}\alpha ^{m}\frac{d_{m}}{m}]\frac{J_{n}}{z^{\alpha n}} 
\]

Thus, in this way we have shown that real order reduced Bessel functions are
deformations of integer orders, similarly to $d_{\lambda }$ being
deformation of $d$.Notice that the deformation mechanism in this case
converts an integer into a real, leading to a new function (which is not
that new !) .This is however not too bad , as we succeeded in unifying
Bessel functions of different orders,which so far remained ``disconnected
''because we ignored the converting mechanism.

In the present paper we apply the formalism to Hermite polynomials $%
H_{n}(z)\ n\in N$ \footnote{%
For a review of Hermite polynomials and table of integral and series see for
instance \cite{chamseddine} \cite{ mekhfi}} and get instead polynomials
which, henceforth, we will call $M,C,$ and $W$ and will denote respectively $%
M_{n\ \alpha ,H}^{s}(z),C_{n\alpha ,H}^{s}(z)\ $and $W_{n\alpha ,H}^{s}(z)\ $%
with $s,n\ \in N\ ,\alpha =\pm 1$ and where $H$ stands for Hermite
polynomials .This notation,specifying the ``root'' polynomial which in this
case is $H$ may seem heavy at first side .It is however justified by our
expectation that not only Hermite polynomials ,but some other orthogonal
polynomials $P_{n}(z)$ , will presumably fit into the scheme and hence will
be deformed in the same way as $H_{n}(z)$. For this reason we anticipate a
common compact notation for all these deformations to come and write at once
the sequence of polynomials as follows: $P_{n}(z)\rightarrow M_{n\alpha
,P}^{s}(z)\Rightarrow C_{n\alpha ,P}^{s}(z)\rightarrow W_{n\alpha
,P}^{s}(z)\Rightarrow P_{n}(z)$ ,but for the rest of the paper ,where only
Hermite polynomials are involved ,we simplify the notation by erasing the
index $H$ .

It is worth to note that the deformation of polynomials , will not convert
integers into reals as for Bessel functions,as this requires necessarily the
presence of the set Z of integers.In the present case only positive integers
are at works i.e. $n\in N.$ To have an idea how it works in the case of $Z$
, let us recall a simple example from \cite{wissale}.Let the generating
function $\Phi (\theta )$ be a periodic function of $\theta $ ( no
dependence on $z$ and $t=e^{i\theta }\ $) and let the operator $\Pi _{m}$
acts on the components $\phi _{n}=\int_{-\pi }^{\pi }e^{in\theta }\Phi
(\theta )d\theta /2\pi $ as $\Pi _{m}\phi _{n}=\phi _{n+m}.$ Then the new
function is

\begin{eqnarray*}
&&\exp [-\lambda \sum_{m\in Z/0}(-1)^{m}\frac{\Pi _{m}}{m}]\phi _{n} \\
&=&\int_{-\pi }^{\pi }\exp [-\lambda \sum_{m\in Z/0}(-1)^{m}\frac{%
e^{im\theta }}{m}]e^{in\theta }\Phi (\theta )d\theta /2\pi \\
&=&\int_{-\pi }^{\pi }e^{i(n+\lambda )}\Phi (\theta )d\theta /2\pi \\
&=&\phi _{n+\lambda }
\end{eqnarray*}

where we applied the known formula $\sum_{m=1}^{m=\infty }(-1)^{m}\frac{\sin
(m\theta )}{m}=-\theta /2$ with $-\pi \langle \theta \langle \pi .$

Bessel functions of real orders $J_{n+\lambda }(z)$ and of integer orders $%
J_{n}(z)$ which are deformations of each other , play separate roles in the
dynamics of physical systems .Roughly speaking real orders may describe
interacting systems whose free parts are described by integer orders .We
have an explicit example of this in topological quantum mechanics \cite
{nadia} .Motivated by this , we expect that such deformation when applied to
Hermite polynomials ( and presumably other orthogonal polynomial describing
a physical system ) will lead ,as for Bessel functions ,to new polynomials
,in the present case , $M_{n\alpha ,H}^{s}(z),C_{n\alpha
,h}^{s}(z),W_{n\alpha ,H}^{s}(z)$ and these will play a role in interacting
systems whose free parts are described by Hermite polynomials.

\section{$M_{n\alpha }^{s}(z)$ Polynomials}

\subsection{Definitions}

The deformation applied to Hermite polynomials $H_{n}(z)$ yields new
polynomials $M_{n\alpha }^{s}(z)$

\begin{equation}
M_{n\alpha }^{s}(z)=\exp [s\sum_{m=1}^{\infty }\alpha ^{m}\frac{d_{m}}{m}]\
H_{n}(z)  \label{eq:7}
\end{equation}

where here $d_{m}$ stands for the m-fold product $d_{m}=\frac{d}{dz}.\,\frac{%
d}{dz}\cdots \cdots \frac{d}{dz}$ .The index $n\in N\ $by definition of
Hermite polynomials while $s$ and $\alpha $ will be fixed by orthogonality
requirements to positive integer values $s\in N$ and $\alpha =\pm 1.$
Hermite polynomials are polynomials of degree n in the z variable.According
to the definition $M_{n\alpha }^{s}(z)$ are also polynomials of degree n in
the z variable .In addition , they are polynomials of degree n in the s
variable too.Note also that according to the recursion property

\begin{equation}
\frac{d}{dz}H_{n}(z)=2nH_{n-1}(z)  \label{eq:7'}
\end{equation}
the series in \ref{eq:7} is finite i.e. we may replace it by $\sum_{m=1}^{n}$%
.

We could have defined $M$ polynomials by their generating function from the
start and get equation \ref{eq:7} afterwards but it is more pedagogical to
start from equation \ref{eq:7} and find the generating function .To find the
generating function we compute the following series \footnote{%
Unless necessary we will use the compact notation $\exp (s\sum_{\alpha })$
to mean $\exp [s\sum_{m=1}^{\infty }\alpha ^{m}\frac{d_{m}}{m}]$}

\begin{eqnarray}
\sum_{n=0}^{\infty }\frac{M_{n\alpha }^{s}(z)}{n!}t^{n}
&=&\sum_{n=0}^{\infty }\exp (s\sum_{\alpha })\frac{H_{n}}{n!}t^{n}
\label{eq:8} \\
&=&\exp (s\sum_{\alpha })\sum_{n=0}^{\infty }\frac{H_{n}}{n!}t^{n}  \nonumber
\\
&=&\exp (s\sum_{\alpha })\exp [-t^{2}+2tz]  \nonumber \\
&=&\exp [s\sum_{m=1}^{\infty }\frac{(2\alpha t)^{m}}{m}]\exp (-t^{2}+2tz) 
\nonumber \\
&=&\frac{\exp (-t^{2}+2tz)}{(1-2\alpha t)^{s}}  \nonumber
\end{eqnarray}

Inverting the series in \ref{eq:8} we get $M$ in an integral form

\[
\frac{M_{n\alpha }^{s}(z)}{n!}=\frac{1}{2\pi i}\oint \frac{\exp (-t^{2}+2tz)%
}{(1-2\alpha t)^{s}t^{n+1}}dt 
\]

If we develop the factor in the denominator

\[
(1-2\alpha t)^{-s}=1+2\alpha st+\frac{s(s+1)}{2!}(2\alpha t)^{2}+\cdots 
\frac{s(s+1)\cdots (s+k-1)}{k!}(2\alpha t)^{k}+\cdots 
\]

we get a useful expansion of $M$ in the index $s$ 
\begin{eqnarray}
M_{n\alpha }^{s}(z) &=&H_{n}(z)+2\alpha s\left( 
\begin{array}{c}
n \\ 
1
\end{array}
\right) H_{n-1}(z)\cdots  \label{eq:11} \\
&&(2\alpha )^{k}s(s+1)\cdots (s+k-1)\left( 
\begin{array}{c}
n \\ 
k
\end{array}
\right) H_{n-k}(z)  \nonumber \\
&&\cdots (2\alpha )^{n}s(s+1)\cdots (s+n-1)  \nonumber
\end{eqnarray}

This expansion shows explicitly that $M_{n\alpha }^{s}(z)$ is a polynomial
of degree n in the $z,s$ variables .To give the polynomials as a power
expansion in the z variable ,we use Hermite polynomial expansion

\begin{equation}
H_{n}(z)=2^{n}\left( 
\begin{array}{c}
n \\ 
0
\end{array}
\right) z^{n}-2^{n-1}\left( 
\begin{array}{c}
n \\ 
2
\end{array}
\right) z^{n-1}+\cdots  \label{eq:12}
\end{equation}

In we plug \ref{eq:12} into \ref{eq:11} we find that $M$ can be organized as
follows 
\begin{equation}
M_{n\alpha }^{s}(z)=\sum_{p=0}^{n}2^{p}\left( 
\begin{array}{c}
n \\ 
n-p
\end{array}
\right) z^{p}M_{(n-p)\alpha }^{s}\ (0)  \label{eq:13}
\end{equation}

where using the values of $H_{2n}(0)=(-1)^{n}2^{n}(2n-1)!!$ and $%
H_{2n+1}(0)=0$ we get after some combinatorics

\begin{equation}
\text{ }M_{n\alpha }^{s}(0)=n!2^{n}\alpha ^{n}\sum_{k=0}^{\frac{n}{2}+\frac{%
(-1)^{n}-1}{4}}\frac{(-1)^{k}}{2^{2k}k!(n-2k)!}\ s(s+1)(s+2)\ldots (s+n-2k-1)
\label{eq:14}
\end{equation}

Let us give for arbitrary s, explicit expressions for the first $M$
polynomials

\begin{eqnarray*}
M_{0\alpha }^{s}(z) &=&1 \\
M_{1\alpha }^{s}(z) &=&2(z+\alpha s) \\
M_{2\alpha }^{s}(z) &=&4(z^{2}+2\alpha sz+s(s+1)-\frac{1}{2}) \\
M_{3\alpha }^{s}(z) &=&8(z^{3}+3\alpha sz^{2}+(3s(s+1)-\frac{3}{2})z+\alpha
(6s+s(s+1)(s+2)))
\end{eqnarray*}

Terms of higher powers in z are simpler to work out as their coefficients $%
M_{n\alpha }^{s}(0)$ only involve lowest $n$

\begin{eqnarray*}
M_{n\alpha }^{s}(z) &=&2^{n}\left( 
\begin{array}{c}
n \\ 
0
\end{array}
\right) z^{n}+2^{n}\left( 
\begin{array}{c}
n \\ 
1
\end{array}
\right) \alpha s\ z^{n-1} \\
&&-2^{n-1}\left( 
\begin{array}{c}
n \\ 
2
\end{array}
\right) (1-2s\ -2s^{2})\ z^{n-2}+\cdots
\end{eqnarray*}

\subsection{Properties of ${\bf M}_{n\alpha }^{s}{\bf (z)}$}

Among the properties we will study in this subsection ,partial-orthogonality
is the most important .The following are a set of relatively simple
properties of $M$ polynomials

\subsubsection{Property 0:Connection to $\;H_{n}(z)$}

\begin{eqnarray*}
M_{n\alpha }^{s}(z) &=&\exp (s\sum_{\alpha })H_{n}(z) \\
H_{n}(z) &=&\exp (-s\sum_{\alpha })M_{n\alpha }^{s}(z) \\
M_{n\alpha }^{s^{\prime }+s}(z) &=&\exp (s^{\prime }\sum_{\alpha
})M_{n\alpha }^{s}(z)
\end{eqnarray*}

\subsubsection{Property 1:Recursion formula involving the derivative}

\[
\frac{dM_{n\alpha }^{s}(z)}{dz}=\exp (s\sum_{\alpha })\frac{dH_{n}(z)}{dz}%
=2n\exp (s\sum_{\alpha })H_{n-1}(z)=2nM_{(n-1)\alpha }^{s}(z) 
\]

where we use the the commutation relation $\exp (s\sum_{\alpha })\frac{d}{dz}%
=\frac{d}{dz}\exp (s\sum_{\alpha })$ together with the Hermite polynomial
property \ref{eq:7'}

\subsubsection{Property 2:Recursion formula in s}

Let us rewrite $M_{n\alpha }^{s^{\prime }}$ in terms of $M_{n\alpha }^{s}$
with $s^{^{\prime }}\geq s$

\begin{eqnarray}
M_{n\alpha }^{s^{\prime }} &=&\exp (s\sum_{\alpha })\exp [(s^{\prime
}-s)\sum_{\alpha }]H_{n}  \label{eq:19} \\
&=&\exp (s\sum_{\alpha })M_{n\alpha }^{s^{\prime }-s}  \nonumber \\
&=&M_{n\alpha }^{s}(z)+2\alpha (s^{\prime }-s)\left( 
\begin{array}{c}
n \\ 
1
\end{array}
\right) M_{(n-1)\alpha }^{s}(z)\cdots  \nonumber \\
&&(2\alpha )^{k}(s^{\prime }-s)(s^{\prime }-s+1)\cdots (s^{\prime
}-s+k-1)\left( 
\begin{array}{c}
n \\ 
k
\end{array}
\right) M_{(n-k)\alpha }^{s}(z) \\
&&\cdots (2\alpha )^{n}(s^{\prime }-s)(s^{\prime }-s+1)\cdots (s^{\prime
}-s+n-1)
\end{eqnarray}

To come to the last line we use \ref{eq:11} and \ref{eq:7}

The particular case $s^{\prime }=s+1$ will be used to write down a recursive
formula for the integral of $M_{n\alpha }^{s^{\prime }}$ . In this
particular case the recursion formula simplifies to the form

\begin{eqnarray*}
M_{n\alpha }^{s+1}(z) &=&M_{n\alpha }^{s}(z)+2\alpha n\ M_{(n-1)\alpha
}^{s}(z)\cdots \\
&&(2\alpha )^{k}\frac{n!}{(n-k)!}M_{(n-k)\alpha }^{s}(z) \\
&&\cdots (2\alpha )^{n}n! \\
&=&\sum_{k=0}^{n}(2\alpha )^{k}\frac{n!}{(n-k)!}M_{(n-k)\alpha }^{s}(z)
\end{eqnarray*}

\subsubsection{Property 3:Recursion formula in n}

Starting from

\begin{equation}
H_{n+1}(z)=2zH_{n}(z)-2nH_{n-1}(z)  \label{eq:24}
\end{equation}

we apply the operator $\exp (s\sum_{\alpha })$ at both sides and get

\begin{eqnarray*}
\exp (s\sum_{\alpha })H_{n+1}(z) &=&2\exp (s\sum_{\alpha
})(zH_{n}(z))-2n\exp (s\sum_{\alpha })H_{n-1}(z) \\
M_{(n+1)\alpha }^{s}(z) &=&2zM_{n\alpha }^{s}-2nM_{(n-1)\alpha }^{s}+2[\exp
(s\sum_{\alpha }),z]H_{n} \\
&=&2zM_{n\alpha }^{s}-2nM_{(n-1)\alpha }^{s}+2s\sum_{m=0}^{n}\alpha
^{m+1}d_{m}M_{n\alpha }^{s} \\
&=&2zM_{n\alpha }^{s}-2nM_{(n-1)\alpha }^{s}+s\sum_{m=0}^{n}(2\alpha )^{m+1}%
\frac{n!}{(n-m)!}M_{(n-m)\alpha }^{s}
\end{eqnarray*}

In line three we used the following result 
\[
\lbrack \exp (s\sum_{\alpha }),z]=s\sum_{m=0}^{\infty }\alpha
^{m+1}d_{m}\exp (s\sum_{\alpha }) 
\]

and applied property 1 to express the m$^{th}$ order derivative of $M$ as $%
d_{m}M_{n\alpha }^{s}=2n2(n-1)\cdots 2(n-m+1)\ M_{(n-m)\alpha }^{s}=2^{m}%
\frac{n!}{(n-m)!}M_{(n-m)\alpha }^{s}$ .

\subsubsection{Property 4 :Partial-orthogonality}

We want to investigate the following integral

\begin{equation}
\int_{-\infty }^{\infty }M_{n\alpha }^{s}(z)\ M_{m\alpha }^{s}(z)\ {\cal D}%
_{s\alpha }(z)\ dz  \label{eq:25}
\end{equation}

The purpose is to see to which extent we can generalize the orthogonality
relation for Hermite polynomials

\begin{equation}
\int_{-\infty }^{\infty }H_{n}(z)\ H_{m}(z)\ \frac{\exp (-z^{2})}{\sqrt{\pi }%
}\ dz\sim \delta _{nm}  \label{eq:25'}
\end{equation}
The function  ${\cal D}_{s\alpha }(z)$ is what generalizes the measure  $%
{\cal D}_{0\alpha }(z)=\frac{1}{\sqrt{\pi }}\exp (-z^{2}).$ We will use
improperly the word measure and orthogonality relations to refer
respectively to  ${\cal D}_{s\alpha }(z)$ and to the orthogonality with
respect to it .  ${\cal D}_{s\alpha }(z)$ is rather a density of a non
definite sign , as we see next.To derive the measure ,we multiply both sides
of \ref{eq:8} by $M_{0\alpha }^{s}(z)=H_{0}(z)=1$ and anticipate
orthogonality of all polynomials $M_{n\alpha }^{s}(z)$ $,$ $n\neq 0$ with $%
M_{0\alpha }^{s}(z)$ i.e.

\[
\int_{-\infty }^{\infty }M_{n\alpha }^{s}(z)\ \ {\cal D}_{s\alpha }(z)\
dz=\delta _{n0}\quad 
\]
This is what we mean by partial-orthogonality .In the case of classic
orthogonal polynomials $P_{n..}$ partial-orthogonality ensures full
orthogonality and vice-versa .i.e. 
\[
\int_{-\infty }^{\infty }P_{n..}(z)\ \ {\cal D}_{..}(z)\ dz\sim \delta
_{n0}\Leftrightarrow \int_{-\infty }^{\infty }P_{n..}(z)\ P_{m..}(z)\ {\cal D%
}_{..}(z)\ dz\sim \delta _{nm}
\]

We will check in the following , that this is not the case for $M$
polynomials .We will however prove the existence of polynomials we denote $%
C_{n\alpha }^{s}(z),$ which for lack of being orthogonal have
``orthogonality'' integrals giving the total charge carried by the field $%
C_{n\alpha }^{s}(z)$ .

In going from $M$ to $C$ in the process of orthogonalization , $C$ appears
to be linear combinations of $M^{^{\prime }s}$ .This linear transformation
is a second map .With two commuting maps (the linear and the exponential )
at hand one can reach polynomials $C$ through two equivalent paths and this
leads us to define a third family of polynomials which we will denote $%
W_{n\alpha }^{s}$ $(z).$ We thus will have the following sequence of
polynomials, all associated with Hermite polynomials which we may draw as $%
H_{n}(z)\rightarrow M_{n\alpha }^{s}(z)\Rightarrow C_{n\alpha
}^{s}(z)\rightarrow W_{n\alpha }^{s}(z)\Rightarrow H_{n}(z)$ or

\begin{equation}
\begin{array}{cc}
\begin{array}{c}
M_{n\alpha }^{s}(z) \\ 
\Downarrow
\end{array}
\leftarrow \!\!\! & 
\begin{array}{c}
H_{n}(z) \\ 
\Uparrow
\end{array}
\\ 
\begin{array}{c}
C_{n\alpha }^{s}(z)
\end{array}
\rightarrow & 
\begin{array}{c}
W_{n\alpha }^{s}(z)
\end{array}
\end{array}
\label{eq:square}
\end{equation}

where $\rightarrow $ and $\Rightarrow $ refer to the exponential and the
linear maps respectively .We will develop all this in the remaining
sections.Coming back to the definition of the measure, the later should
satisfy the equation

\begin{equation}
\int_{-\infty }^{\infty }\frac{\exp (-t^{2}+2tz)}{(1-2\alpha t)^{s}}{\cal D}%
_{s\alpha }(z)dz=1  \label{eq:27}
\end{equation}

We therefore have to choose the measure  ${\cal D}_{s\alpha }(z)$ such that
it absorbs the dependence on $t.\func{Re}$stricting $s$ to $s\in N$ the
following form of the measure is a solution to \ref{eq:27}

\begin{equation}
{\cal D}_{s\alpha }(z)=\frac{\alpha ^{s}}{\sqrt{\pi }}\exp (-\alpha z)\frac{%
d^{s}}{dz^{s}}\exp (-z^{2}+\alpha z)  \label{eq:28}
\end{equation}

In fact inserting \ref{eq:28} into \ref{eq:27} we get

\begin{eqnarray*}
&&\int_{-\infty }^{\infty }\frac{\exp (-t^{2}+2tz)}{(1-2\alpha t)^{s}}\frac{%
\alpha ^{s}}{\sqrt{\pi }}\exp (-\alpha z)\frac{d^{s}}{dz^{s}}\exp
(-z^{2}+\alpha z)dz \\
&=&\frac{(-\alpha )^{s}}{\sqrt{\pi }}\int_{-\infty }^{\infty }\frac{\exp
(-z^{2}+\alpha z)}{(1-2\alpha t)^{s}}\frac{d^{s}}{dz^{s}}\exp
(-t^{2}+z(-\alpha +2t))dz \\
&=&\frac{(-\alpha )^{s}}{\sqrt{\pi }}\left( \frac{-\alpha +2t}{1-2\alpha t}%
\right) ^{s}\int_{-\infty }^{\infty }\exp (-(z-t)^{2})dz \\
&=&\left( \frac{\alpha -2t}{\frac{1}{\alpha }-2t}\right) ^{s}
\end{eqnarray*}

In the second line an integration by part has been performed .The last line
shows that the form of the measure we proposed in \ref{eq:28} is the
appropriate measure provided $s\in N$ and $\alpha =\pm 1.$ Let us rewrite
the measure in a more practical form (Hermite function)

\begin{eqnarray*}
{\cal D}_{s\alpha }(z) &=&\frac{\alpha ^{s}}{\sqrt{\pi }}\exp (-\alpha z)%
\frac{d^{s}}{dz^{s}}\exp (-z^{2}+\alpha z) \\
&=&\frac{(-\alpha )^{s}}{\sqrt{\pi }}\exp (-z^{2})H_{s}(z-\frac{\alpha }{2})
\end{eqnarray*}

where we used the standard form of Hermite polynomials $(-1)^{s}\exp (z^{2})%
\frac{d^{s}}{dz^{s}}\exp (-z^{2}).$

It is worth to note ,at this point ,that the function  ${\cal D}_{s\alpha
}(z)$ is not a positive real function.It is rather a density with a finite
positive total ``charge''

\begin{equation}
\int_{-\infty }^{\infty }{\cal D}_{s\alpha }(z)dz=(\alpha ^{2})^{s}=1
\label{eq:17}
\end{equation}

The orthogonality integral $\int_{-\infty }^{\infty }C_{n\alpha
}^{s}(z)C_{m\alpha }^{s}(z)$  ${\cal D}_{s\alpha }(z)dz\sim \delta _{nm}$
which generalizes \ref{eq:25'} will have the meaning of the charge carried
by the field $C_{n\alpha }^{s}(z)$ .The waves will have positive charges
(positive sign) ,negative charges (negative sign) or zero charges (vanishing
integral),contrary to the orthogonality integral of Hermite polynomials
which in this physical language ,describes the probability to find
everywhere the system described by $H_{n}$ .This probability is proportional
to $\int_{-\infty }^{\infty }H_{n}^{2}(z)\exp (-z^{2})dz\ $and is therefore
strictly positive.

Now we proceed to work out a recursion formula in the index $s$ for the
integral \ref{eq:25}.Integrals of $M$ are the building blocks of polynomials 
$C$ as we will see in the next section .In practice the iterative procedure
will be very useful as it will allow to handle these integrals using
computer programming .We thus have to relate ${\cal I}_{(nm)\alpha }^{s+1} $
to ${\cal I}_{(nm)\alpha }^{p};$ $p=0,1,\cdots s$ where the integral is that
defined in \ref{eq:25}

\[
{\cal I}_{(nm)\alpha }^{s+1}=\int_{-\infty }^{\infty }M_{n\alpha }^{s+1}(z)\
M_{m\alpha }^{s+1}(z)\ \frac{(-\alpha )^{s+1}}{\sqrt{\pi }}\exp
(-z^{2})H_{s+1}(z-\frac{\alpha }{2})\ dz 
\]

Expanding $M_{n\alpha }^{s+1}$ in terms of $M_{n\alpha }^{s}$ by using
property 2 applied to $s^{^{\prime }}=s+1$we get

\[
{\cal I}_{(nm)\alpha }^{s+1}=\int_{-\infty }^{\infty
}\sum_{k=0}^{n}\sum_{l=0}^{m}a_{k\alpha }^{n}a_{l\alpha }^{m}M_{(n-k)\alpha
}^{s}(z)\ M_{(m-l)\alpha }^{s}(z)\frac{\ (-\alpha )^{s+1}}{\sqrt{\pi }}\exp
(-z^{2})H_{s+1}(z-\frac{\alpha }{2})\ dz 
\]

with $a_{m\alpha }^{n}=(2\alpha )^{m}\frac{n!}{(n-m)!},$then rewriting the
Hermite polynomial $H_{s+1}(z-\frac{\alpha }{2})$ using the recursion
formulas \ref{eq:7'} and \ref{eq:24} as

\begin{equation}
H_{s+1}(z-\frac{\alpha }{2})=(2z-\alpha )H_{s}(z-\frac{\alpha }{2})-\frac{%
dH_{s}(z-\frac{\alpha }{2})}{dz}  \label{eq:33}
\end{equation}

we consider the second term in \ref{eq:33} and perform an integration by
part( the factor $e^{-z^{2}}$ suppresses boundary terms)

\begin{eqnarray*}
&&-\int_{-\infty }^{\infty }\sum_{k=0}^{n}\sum_{l=0}^{m}a_{k\alpha
}^{n}a_{l\alpha }^{m}M_{z(n-k)\alpha }^{s}(z)\ M_{(m-l)\alpha }^{s}(z)\ 
\frac{(-\alpha )^{s+1}}{\sqrt{\pi }}e^{-z^{2}}\frac{dH_{s}(z-\frac{\alpha }{2%
})}{dz}dz \\
&=&-\alpha \int_{-\infty }^{\infty }\sum_{k=0}^{n}\sum_{l=0}^{m}a_{k\alpha
}^{n}a_{l\alpha }^{m}\frac{d}{dz}\left( e^{-z^{2}}M_{z(n-k)\alpha }^{s}(z)\
M_{(m-l)\alpha }^{s}(z)\ \right) \frac{(-\alpha )^{s}}{\sqrt{\pi }}H_{s}(z-%
\frac{\alpha }{2})dz \\
&=&\alpha \int_{-\infty }^{\infty }\sum_{k=0}^{n}\sum_{l=0}^{m}a_{k\alpha
}^{n}a_{l\alpha }^{m}e^{-z^{2}} \\
&&(2z\ M_{(n-k)\alpha }^{s}(z)\ M_{(m-l)\alpha }^{s}(z)\  \\
&&-2(n-k)\ M_{(n-k-1)\alpha }^{s}(z)\ M_{(m-l)\alpha }^{s}(z) \\
&&-2(m-l)\ M_{(n-k)\alpha }^{s}(z)\ M_{(m-l-1)\alpha }^{s}(z)) \\
&&\frac{(-\alpha )^{s}}{\sqrt{\pi }}H_{s}(z-\frac{\alpha }{2})dz
\end{eqnarray*}

In performing differentiation with respect to z we used property 1 and $%
\alpha ^{2}=\pm 1$.Reporting the first term in \ref{eq:33}

\[
-\alpha \int_{-\infty }^{\infty }\sum_{k=0}^{n}\sum_{l=0}^{m}a_{k\alpha
}^{n}a_{l\alpha }^{m}M_{(n-k)\alpha }^{s}(z)\ M_{(m-l)\alpha }^{s}(z)\ \frac{%
(-\alpha )^{s}}{\sqrt{\pi }}\exp (-z^{2})(2z-\alpha )H_{s}(z-\frac{\alpha }{2%
})dz 
\]

we see that terms ``linear ''in $z$ simplify , leading to the expression

\begin{eqnarray*}
{\cal I}_{(nm)\alpha }^{s+1} &=&\int_{-\infty }^{\infty
}\sum_{k=0}^{n}\sum_{l=0}^{m}(a_{k\alpha }^{n}a_{l\alpha }^{m}M_{(n-k)\alpha
}^{s}(z)\ M_{(m-l)\alpha }^{s}(z)\  \\
&&-2(n-k)\ \alpha a_{k\alpha }^{n}a_{l\alpha }^{m}M_{(n-k-1)\alpha }^{s}(z)\
M_{(m-l)\alpha }^{s}(z) \\
&&-2(m-l)\ \alpha a_{k\alpha }^{n}a_{l\alpha }^{m}M_{(n-k)\alpha }^{s}(z)\
M_{(m-l-1)\alpha }^{s}(z)) \\
&&\frac{(-\alpha )^{s}}{\sqrt{\pi }}\exp (-z^{2})H_{s}(z-\frac{\alpha }{2})\
dz
\end{eqnarray*}

Then we use the self-evident relation $2(n-p)\alpha a_{p\alpha
}^{n}=\,a_{(p+1)\alpha }^{n}$ ( see the definition of the $a^{^{\prime }s}$
above ) and get

\begin{eqnarray*}
{\cal I}_{(nm)\alpha }^{s+1} &=&\int_{-\infty }^{\infty }M_{n\alpha
}^{s}(z)\ M_{m\alpha }^{s}(z){\cal D}_{s\alpha }(z)dz \\
&&+\int_{-\infty }^{\infty }(\left(
\sum_{k=0}^{n}\sum_{l=1}^{m}+\sum_{k=1}^{n}\sum_{l=0}^{m}-\sum_{k=1}^{n}%
\sum_{l=1}^{m}\right) (a_{k\alpha }^{n}a_{l\alpha }^{m}M_{(n-k)\alpha
}^{s}(z)\ M_{(m-l)\alpha }^{s}(z) \\
&&-\sum_{k=0}^{n}\sum_{l=0}^{m}a_{(k+1)\alpha }^{n}a_{l\alpha
}^{m}M_{(n-k-1)\alpha }^{s}(z)\ M_{(m-l)\alpha }^{s}(z) \\
&&-\sum_{k=0}^{n}\sum_{l=0}^{m}a_{k\alpha }^{n}a_{(l+1)\alpha
}^{m}M_{(n-k)\alpha }^{s}(z)\ M_{(m-l-1)\alpha }^{s}(z)) \\
&&{\cal D}_{s\alpha }(z)dz
\end{eqnarray*}
In the first and the second line we made the decomposition $%
\sum_{k=0}^{n}\sum_{l=0}^{m}$ $a_{k\alpha }^{n}a_{l\alpha }^{m}\cdots =$ $%
1+(\sum_{k=0}^{n}\sum_{l=1}^{m}+\sum_{k=1}^{n}\sum_{l=0}^{m}-\sum_{k=1}^{n}%
\sum_{l=1}^{m})$ $a_{k\alpha }^{n}a_{l\alpha }^{m}$ $\cdots .$ To continue
we make the substitution $k\rightarrow k-1$ in the third term and $%
l\rightarrow l-1$ in the fourth term and rewrite ${\cal I}_{(nm)\alpha
}^{s+1}$ as

\begin{eqnarray*}
{\cal I}_{(nm)\alpha }^{s+1} &=&{\cal I}_{(nm)\alpha }^{s} \\
&&+\left(
\sum_{k=0}^{n}\sum_{l=1}^{m}+\sum_{k=1}^{n}\sum_{l=0}^{m}-\sum_{k=0}^{n+1}%
\sum_{l=1}^{m}-\sum_{k=1}^{n}\sum_{l=0}^{m+1}-\sum_{k=1}^{n}\sum_{l=1}^{m}%
\right) a_{k\alpha }^{n}a_{l\alpha }^{m}M_{(n-k)\alpha }^{s}(z)\
M_{(m-l)\alpha }^{s}(z)\  \\
&&\frac{(-\alpha )^{s}}{\sqrt{\pi }}\exp (-z^{2})H_{s}(z-\frac{\alpha }{2})\
dz
\end{eqnarray*}

The first four $\sum $ terms cancel each other as $k$ and $l$ cannot exceed
the values $m$ and $n$ respectively ( $M$ polynomials as Hermite polynomials
vanish identically for negative indices).We thus end up with the
partial-orthogonality formula (which if full orthogonality was realized
reduces to the first term)

\begin{eqnarray*}
{\cal I}_{(nm)\alpha }^{s+1} &=&{\cal I}_{(nm)\alpha
}^{s}-\sum_{k=1}^{n}\sum_{l=1}^{m}a_{k\alpha }^{n}a_{l\alpha }^{m}{\cal I}%
_{(nm)\alpha }^{s} \\
&=&{\cal I}_{(nm)\alpha
}^{0}.-\sum_{k=1}^{n}\sum_{l=1}^{m}\sum_{p=0}^{s}a_{k\alpha }^{n}a_{l\alpha
}^{m}{\cal I}_{(nm)\alpha }^{p} \\
&=&2^{n}n!\delta _{nm}-\sum_{k=1}^{n}\sum_{l=1}^{m}\sum_{p=0}^{s}a_{k\alpha
}^{n}a_{l\alpha }^{m}{\cal I}_{(nm)\alpha }^{p}
\end{eqnarray*}

Let us give some explicit calculations using the above formula

\begin{eqnarray*}
{\cal I}_{(n1)\alpha }^{s} &=&-(2\alpha )^{n+1}n!s\quad ;\quad n\neq 0,1 \\
{\cal I}_{(22)\alpha }^{s} &=&16(2s^{2}-8s+\frac{1}{2}) \\
{\cal I}_{(32)\alpha }^{s} &=&384s(s-\frac{5}{2})\alpha
\end{eqnarray*}

\subsection{Differential equation}

Hermite polynomials satisfy a second order ordinary differential equation

\begin{equation}
(\frac{d^{2}}{dz^{2}}-2z\frac{d}{dz}+2n)H_{n}(z)=0  \label{eq:45}
\end{equation}

To find the differential equation satisfied by $M$, the best way is to map 
\ref{eq:45} by the operator $\exp $(s$\sum ).$ Knowing that this operator,
by construction, commutes with $\frac{d^{2}}{dz^{2}}$ and$\frac{d}{dz}$ we
get the following equation

\begin{eqnarray}
&&(\frac{d^{2}}{dz^{2}}-2z\frac{d}{dz}+2n)M_{n\alpha }^{s}(z)  \label{eq:45"}
\\
&=&2\left[ \exp (s\sum ),z\right] \frac{d}{dz}H_{n}(z)  \nonumber \\
&=&2s\sum_{m=1}^{\infty }\alpha ^{m}\frac{d^{m}}{dz^{m}}M_{n\alpha }^{s}(z) 
\nonumber \\
&=&2s(2\alpha )^{n}\sum_{p=0}^{n-1}\frac{n!}{(2\alpha )^{p}p!}M_{p\alpha
}^{s}  \nonumber
\end{eqnarray}

In the third line we inserted the commutator used earlier

$[\exp (s\sum_{\alpha }),z]=s\sum_{m=1}^{\infty }\alpha ^{m}\frac{d^{m-1}}{%
dz^{m-1}}\exp (s\sum_{\alpha })$ ( which has the appropriate form to
reproduce a differential equation) and in the fourth line we used property 1
to write $\sum_{m=1}^{n}\alpha ^{m}\frac{d^{m}}{dz^{m}}M_{n\alpha
}^{s}(z)=\sum_{p=0}^{n-1}(2\alpha )^{n-p}\frac{n!}{p!}M_{p\alpha }^{s}$ Note
as before that the series in the above equation is finite( $1\leq m\leq n$
when applied to $M_{n}^{s}\alpha $ )as $M^{^{\prime }s}$ are identically
vanishing for negative $n$ and each derivative lowers the index by one unit
according to property 1.

This is a differential equation we can interpret equivalently as an equation
of order $n$ with varying coefficients or a linear system of second order
differential equations or ,if we plug the explicit expression for the
polynomials $M$ into the $\sum $ , as an inhomogeneous differential equation
whose homogeneous part is the differential equation of Laguerre polynomials.

The linear system has the form

\[
\left| 
\begin{array}{llllllllll}
D &  &  &  &  & \cdots \cdots 0 & \cdots  & \cdots  & \cdots  & 0 \\ 
& D+2 &  &  &  &  & \ddots  &  &  & \vdots  \\ 
&  & \vdots  & \ddots  &  &  &  & \ddots  &  & \vdots  \\ 
\cdots  &  & -2s(2\alpha )^{j-i}\frac{j!}{i!} & \cdots  &  & D+2j &  &  & 
\ddots  & \vdots  \\ 
&  & \vdots  &  &  &  &  &  &  & 0 \\ 
& \cdots  & -2s(2\alpha )^{n-i}\frac{n!}{i!} & \cdots  &  &  &  &  &  & D+2n
\end{array}
\right| \left| 
\begin{array}{c}
M_{0\alpha }^{s} \\ 
M_{1\alpha }^{s} \\ 
\vdots  \\ 
\vdots  \\ 
\vdots  \\ 
M_{n\alpha }^{s}
\end{array}
\right| =0
\]

where $D=\frac{d^{2}}{dz^{2}}-2z\frac{d}{dz}$ and the matrix is an $%
(n+1)\times (n+1)$ triangular matrix.

Plugging ,for illustration, only the higher powers of the series in the last
line of \ref{eq:45"}

\[
\sum_{p=0}^{n-1}\frac{M_{p\alpha }^{s}(z)}{(2\alpha )^{p}p!}=\frac{(\alpha
z)^{n-1}}{(n-1)!}+\frac{(\alpha z)^{n-2}}{(n-2)!}(1+s)+\cdots 
\]

we get the inhomogeneous differential equation

\[
(\frac{d^{2}}{dz^{2}}-2z\frac{d}{dz}+2n)M_{n\alpha }^{s}(z)=s2^{n+1}(n\alpha
\ z^{n-1}+n(n-1)(1+s)\ z^{n-2}+\cdots \cdots ) 
\]

\section{$C_{n\alpha }^{s}(z)$ $,W_{n\alpha }^{s}(z)$ Polynomials}

\subsection{Introduction}

In the previous section we defined the measure to be

\[
{\cal D}_{s\alpha }(z)=\frac{(-\alpha )^{s}}{\sqrt{\pi }}\exp
(-z^{2})H_{s}(z-\frac{\alpha }{2})
\]

Associated with this function is the following integral

\begin{equation}
\int_{-\infty }^{\infty }z^{n}{\cal D}_{s\alpha }(z)dz  \label{eq:48'}
\end{equation}

for $n=1,2,$ $\ldots .$ .which may be shown to be finite .In fact we can
compute it for any value $n$, in principle.To proceed we first express the
quantity $z^{n}$  ${\cal D}_{s\alpha }(z)$ as a linear combination of  $%
{\cal D}_{p\alpha }(z)$ with $(s-n)\leq p\leq (s+n).$ i.e.

\begin{equation}
z^{n}{\cal D}_{s\alpha }(z)=\sum_{(s-n)\leq p\leq (s+n)}d_{p\alpha }^{sn}%
{\cal D}_{p\alpha }(z)  \label{eq:49}
\end{equation}

This is the general pattern as we can see from first components computations.

\begin{eqnarray*}
\frac{z}{\alpha }{\cal D}_{s\alpha }(z) &=&-s{\cal D}_{(s-1)\alpha }(z)+%
\frac{1}{2}{\cal D}_{s\alpha }(z)-\frac{1}{2}{\cal D}_{(s+1)\alpha }(z) \\
z^{2}{\cal D}_{s\alpha }(z) &=&s(s-1){\cal D}_{(s-2)\alpha }(z)-s{\cal D}%
_{(s-1)\alpha }(z) \\
&&+(s+\frac{3}{4}){\cal D}_{s\alpha }(z)-\frac{1}{2}{\cal D}_{(s+1)\alpha
}(z)+\frac{1}{4}{\cal D}_{(s+2)\alpha }(z)
\end{eqnarray*}

Let us work out explicitly the first component for illustration

\begin{eqnarray*}
z{\cal D}_{s\alpha }(z) &=&(-\alpha )^{s}\exp (-z^{2})zH_{s}(z-\frac{\alpha 
}{2}) \\
&=&\frac{1}{2}(-\alpha )^{s}\exp (-z^{2})2(z-\frac{\alpha }{2})H_{s}(z-\frac{%
\alpha }{2})+\frac{\alpha }{2}(-\alpha )^{s}\exp (-z^{2})H_{s}(z-\frac{%
\alpha }{2}) \\
&=&\frac{1}{2}(-\alpha )^{s}\exp (-z^{2})\left( H_{s+1}(z-\frac{\alpha }{2}%
)+2sH_{s-1}(z-\frac{\alpha }{2})\right)  \\
&&+\frac{\alpha }{2}(-\alpha )^{s}\exp (-z^{2})H_{s}(z-\frac{\alpha }{2}) \\
&=&-s\alpha (-\alpha )^{s-1}\exp (-z^{2})H_{s-1}(z-\frac{\alpha }{2})+\frac{%
\alpha }{2}(-\alpha )^{s}\exp (-z^{2})H_{s}(z-\frac{\alpha }{2}) \\
&&-\frac{\alpha }{2}(-\alpha )^{s+1}\exp (-z^{2})H_{s+1}(z-\frac{\alpha }{2})
\\
&=&-s\alpha {\cal D}_{(s-1)\alpha }(z)+\frac{\alpha }{2}{\cal D}_{s\alpha
}(z)-\frac{\alpha }{2}{\cal D}_{(s+1)\alpha }(z)
\end{eqnarray*}

With the normalization of  ${\cal D}_{s\alpha }(z)$ such that its integral%
\ref{eq:17} does not depend on the index $s$ and formula \ref{eq:49} the
above integral can be computed trivially

\begin{eqnarray*}
\int_{-\infty }^{\infty }z^{n}{\cal D}_{s\alpha }(z)dz &=&\int_{-\infty
}^{\infty }\sum_{(s-n)\leq p\leq (s+n)}d_{p\alpha }^{s}{\cal D}_{p\alpha }(z)
\\
&=&\sum_{(s-n)\leq p\leq (s+n)}d_{p\alpha }^{s}\int_{-\infty }^{\infty }%
{\cal D}_{p\alpha }(z) \\
&=&\sum_{(s-n)\leq p\leq (s+n)}d_{p\alpha }^{s}
\end{eqnarray*}

Now we know that,for any real function of a real variable z such that the
integral \ref{eq:48'} exists for $n=0,1,\ldots $ ,which condition is
satisfied in the present case,there exists a sequence of polynomials $%
C_{0\alpha }^{s}(z),C_{1\alpha }^{s}(z),\ldots C_{n\alpha }^{s}(z),$ that is
uniquely determined by the following conditions

\begin{enumerate}
\item\begin{itemize}
\item  $C_{n\alpha }^{s}(z)$ is a polynomials of degree $n$

\item  The polynomials $C_{0\alpha }^{s}(z),C_{1\alpha }^{s}(z),\ldots
C_{n\alpha }^{s}(z)$ are orthogonal (The normalization factor $N_{\alpha
}^{s}$ is not necessary positive due to ${\cal D}_{s\alpha }(z)$ being of a
non definite sign ) 
\[
\int_{-\infty }^{\infty }C_{n\alpha }^{s}(z)C_{m\alpha }^{s}(z){\cal D}%
_{s\alpha }(z)dz=N_{n\alpha }^{s}\delta _{nm}
\]
\end{itemize}
\end{enumerate}

\subsection{Definition.}

What relation have polynomials $C$ to existing polynomials ,in this case $H$
and $M$ ?.We previously found that polynomials $M_{n\alpha }^{s}(z)$ are
partial-orthogonal with respect to the measure  ${\cal D}_{s\alpha }(z).$ $%
C_{n\alpha }^{s}(z)$ which are by definition orthogonal with respect to  $%
{\cal D}_{s\alpha }(z)$ are necessarily partial-orthogonal ,hence the most
general form of $C_{n\alpha }^{s}(z)$ ensuring partial-orthogonality is a
linear combination of $M_{n\alpha }^{s}(z)$ \footnote{%
A direct computation of polynomials $C$ using orthogonality with respect to
the measure${\frak \ D}_{s\alpha }(z)$ and expanding $C$ in powers of $z$
shows that the expansion \ref{eq:51} is indeed correct} , $n\neq 0$

\begin{eqnarray}
C_{0\alpha }^{s}(z) &=&M_{0\alpha }^{s}(z)=1  \label{eq:51} \\
C_{n\alpha }^{s}(z) &=&\sum_{i=0}^{n-1}w_{i}^{n}M_{(n-i)\alpha }^{s}(z) 
\nonumber
\end{eqnarray}

where the coefficients $w_{i}^{n}\equiv w_{p\alpha }^{sn}$ are to be
determined using the orthogonality relation .The normalization of the
polynomials above is such that $w_{0}^{n}=1$.

\smallskip Using orthogonality we can work out the coefficients of the above
expansion by iteration as usual.The general pattern is as follows.Define the
determinant $\Delta _{n}$ and $\Delta _{n}^{^{\prime }i}$ where we use the
simpler notation $M_{n}M_{m}\equiv \int_{-\infty }^{\infty }M_{n}(z)M_{m}(z)$
${\cal D}_{s\alpha }(z)dz$ and where $i$ refers to the insertion at the $%
i^{th}$ column $1\leq i\leq n$

\begin{eqnarray*}
\Delta _{n} &=&\left| 
\begin{array}{ccccc}
M_{1}M_{n} & \cdots & \cdots & M_{1}M_{2} & \text{ }M_{1}M_{1} \\ 
M_{2}M_{n} & \cdots & \cdots & M_{2}M_{2} & M_{2}M_{1} \\ 
\vdots & \cdots & \cdots & \vdots & \vdots \\ 
\vdots & \cdots & \cdots & \vdots & \vdots \\ 
M_{n}M_{n} & \cdots & \cdots & M_{n}M_{2} & M_{n}M_{1}
\end{array}
\right| \\
&&\bigskip \\
\Delta _{n}^{^{\prime }i} &=&-\left| 
\begin{array}{cccccc}
M_{1}M_{n} & \cdots & M_{n+1}M_{1} & \cdots & M_{1}M_{2} & \text{ }M_{1}M_{1}
\\ 
M_{2}M_{n} & \cdots & M_{n+1}M_{2} & \cdots & M_{2}M_{2} & M_{2}M_{1} \\ 
\vdots & \cdots &  & \cdots & \vdots & \vdots \\ 
\vdots & \cdots &  & \cdots & \vdots & \vdots \\ 
M_{n}M_{n} & \cdots & M_{n+1}M_{n} & \cdots & M_{n}M_{2} & M_{n}M_{1}
\end{array}
\right|
\end{eqnarray*}

the coefficients are then given by the formula

\[
w_{i}^{n}=\frac{\Delta _{n-1}^{^{\prime }i}}{\Delta _{n-1}} 
\]

The expansion of $C_{n\alpha }^{s}(z)$ in terms of $M_{n\alpha }^{s}(z)$ in
Formula \ref{eq:51} is important as it relates $C_{n\alpha }^{s}(z)$ (
through $M_{n\alpha }^{s}(z)$ ) to the general scheme of Hermite polynomials
deformation using the exponential mapping .In fact we can rewrite \ref{eq:51}
as follows

\begin{eqnarray*}
C_{n\alpha }^{s}(z) &=&\sum_{i=0}^{n-1}w_{i}^{n}M_{(n-i)\alpha }^{s}(z) \\
&=&\exp (s\sum_{\alpha })\sum_{i=0}^{n-1}w_{i}^{n}H_{(n-i)}(z) \\
&=&\exp (s\sum_{\alpha })W_{n\alpha }^{s}(z)
\end{eqnarray*}

Thus $C_{n\alpha }^{s}(z)$ polynomials are deformations of a specific
combination ( linear map ) of Hermite polynomials which we denote $%
W_{n\alpha }^{s}(z)$

\[
W_{n\alpha }^{s}(z)=\sum_{i=0}^{n-1}w_{i}^{n}H_{(n-i)}(z) 
\]
In all we have three families of polynomials $M_{n\alpha }^{s}(z),C_{n\alpha
}^{s}(z),W_{n\alpha }^{s}(z)$ related to Hermite polynomials $H_{n}(z)$
,thus forming a sequence of four families of polynomials as we anticipated
in \ref{eq:square}

\section{Conclusion}

\smallskip In \cite{wissale} we applied a deformation mechanism to Bessel $%
J_{n}(z)$ and Neumann $N_{n}(z)$ functions of integer orders to generate
real order Bessel and Neumann functions respectively $J_{n+\lambda }(z)$ and 
$N_{n+\lambda }(z)$ ($\lambda $ real ) .Here we apply the same deformation
mechanism to Hermite polynomials $H_{n}(z)$ and generate Hermite associated
polynomials $M_{n\beta ,H}^{s}(z);C_{n\beta ,H}^{s}(z);W_{n\beta ,H}^{s}(z)$%
{\bf \ .}The structure underlying such deformation has the following
characteristics.

\begin{itemize}
\item  The associated polynomials come in triplicate and form the sequence

$H_{n}(z)\rightarrow M_{n\alpha ,H}^{s}(z)\Rightarrow C_{n\alpha
,H}^{s}(z)\rightarrow W_{n\alpha ,H}^{s}(z)\Rightarrow H_{n}(z)$

\item  When we try to define a measure ${\cal D}_{s}$ which ensures
orthogonality of polynomials\thinspace \thinspace \thinspace \thinspace $\,M$
$^{^{\prime }s}$ we find that ${\cal D}_{s}$ is not positive .It is rather a
``charge'' density .Moreover this measure has the form ${\cal D}_{s}\sim $ $%
{\cal D}_{0}H_{s}$ where ${\cal D}_{0}$ is the measure of Hermite polynomials

\item  The differential equation associated to $M$ is inhomogeneous whose
homogeneous part is Hermite polynomials differential equation.The
inhomogeneous term being of the form $\sim $ $s\sum_{m=1}^{n}\alpha ^{m}%
\frac{d^{m}}{dz^{m}}M_{n\alpha ,H}^{s}(z)$
\end{itemize}

On the other hand the above structure seems to be general and not restricted
to Hermite polynomials.In fact ,in a parallel ,similar study \cite{mustapha}
, where the deformation is applied to Laguerre polynomials ,the same
underlying structure emerges with the above three characteristics , where $H$
is replaced by $L.$

\end{document}